\RequirePackage{fix-cm}
\documentclass[smallcondensed]{svjour3}     
\smartqed  
\usepackage{graphicx}
\usepackage{amsmath,bm} 
\usepackage{xcolor}
\usepackage{hyperref}
\usepackage[sort, numbers]{natbib}

\begin{document}

\title{Cure models to estimate time until hospitalization due to COVID-19 
}
\subtitle{A case study in Galicia (NW Spain)}


\author{Maria Pedrosa-Laza         \and
        Ana L\'opez-Cheda \and 
				Ricardo Cao
}


\institute{M. Pedrosa-Laza \at
              \'Area de Proyectos de Ingenier\'ia, Escuela T\'ecnica Superior de Minas,  University of Oviedo, Spain \\
              Tel.: +34-985455865\\
              \email{maria.pedrosa@api.uniovi.es}           
           \and
           A. L\'opez-Cheda \at 
              Research Group MODES, CITIC, University of A Coru\~na, 15071 A Coru\~na, Spain \\
							Tel.: +34-881015513\\
							\email{ana.lopez.cheda@udc.es} \\
							\url{http://orcid.org/0000-0002-3618-3246}
						\and
						R. Cao \at  
						Research Group MODES, CITIC, University of A Coru\~na, 15071 A Coru\~na, Spain \\
						and ITMATI \\
						Tel.: +34-881011225\\
						\email{rcao@udc.es} \\
						\url{http://orcid.org/0000-0001-8304-687X}
}

\date{Received: date / Accepted: date}

\maketitle

\begin{abstract}
A short introduction to survival analysis and censored data is included in this paper. A thorough literature review in the field of cure models has been done. An overview on the most important and recent approaches on parametric, semiparametric and nonparametric mixture cure models is also included. The main nonparametric and semiparametric approaches were applied to a real time dataset of COVID-19 patients from the first weeks of the epidemic in Galicia (NW Spain). The aim is to model the elapsed time from diagnosis to hospital admission. The main conclusions, as well as the limitations of both the cure models and the dataset, are presented, illustrating the usefulness of cure models in this kind of studies, where the influence of age and sex on the time to hospital admission is shown. 

\end{abstract}

\section{Introduction}
\label{intro}
\subsection{Survival analysis}
\label{sec:1}
Survival analysis is the branch of Statistics which considers the study of the elapsed time until the occurrence of an event of interest \cite{Singh}. Frequently, such event is death by a pathology, and thus this variable receives the name of ``lifetime", and the event is called ``failure" or ``death". 

\paragraph{}
From a statistical point of view, the survival function at time $t$ is conceived as the probability of an individual living beyond that time. As a result, the basis of survival analysis relies on the estimation of such a probability for any value of $t$. Mathematically, this survival function is defined as
\begin{equation*}
S(t)=P(T>t),
\end{equation*}
where $T$ represents the ``lifetime". It is important to highlight that $S(t)$ can take any shape which satisfies the following conditions  \cite{Klein1}:
\begin{itemize}
\item $S(t)$ is a decreasing function
\item $S(0)=1$ and $\lim_{t \rightarrow \infty} S(t) = 0$
\end{itemize}

A key concept in survival analysis is the hazard function, $h(t)$, which represents the instantaneous failure rate for a certain individual. That is, this function represents the probability that a subject will experience an event of interest within a small specific time interval, given that the individual has survived until the beginning of this interval, defined by:
\begin{equation*}
h(t)=\lim_{\Delta t \rightarrow 0} \frac{P[t \leq T < t + \Delta t | T \geq t]}{\Delta t}.
\end{equation*}
Note that if $T$ is a continuous variable, $S(t)$ and $h(t)$ are directly related through
\begin{equation*}
h(t)=- \frac{d \ln[S(t)]}{dt}.
\end{equation*}

A wide variety of inference techniques can be applied to estimate the survival function and, due to its direct relationship, for the hazard function. The idea is to use samples extracted from a specific population, in order to assess the behavior of the time-to-event variable. Furthermore, these analyses can be useful for the comparison of the survival curves estimated from different populations, or for the assessment of the influence of certain explanatory variables on the lifetime of a group \cite{Kleinbaum}. 

\subsection{Censored data}
\label{sec:2}
There are some limitations regarding data availability when performing a survival study. In some cases, the study design does not allow for the accurate measure of the lifetime for all the individuals within the sample, and this phenomenon is known as censoring. As an illustrative example, let us define the population of ``patients diagnosed with a terminal disease", where the variable of study would be ``elapsed time from the diagnosis until death". Due to specific circumstances that may happen during the follow-up period, such as hospital transfer, leaving the study prior to experiencing the event or premature end of study, the lifetime of some individuals will be unknown. In such cases, these observations are said to be censored. 

\paragraph{}
Censoring plays a key role in survival analysis, and its influence needs to be specifically considered. We can find two major types of censoring \cite{Prinja}:

\begin{itemize}
\item Point censoring, which arises when the event of interest does not occur within the study period, known as right censoring, or when we cannot determine the exact lifetime even if the event of interest happens in between the limits of this period. In the previous example, an observation of a certain individual would be right-censored if death occurs after the end of the study, or if there is a loss of follow-up during the study. On the other hand, the observation will be left-censored if death occurs in the follow-up period but we cannot determine the date of diagnosis, previous to the beginning of the study. 

\item Interval censoring, which appears in these cases when the event happens between two exact time points, but it is not possible to determine the exact point of occurrence within the interval. For example, if the study variable is ``time to recurrence" of a certain disease, the data will be interval censored if a patient does not suffer from the disease in a follow-up visit, but he or she does present it in the following medical check-up, being thus impossible to determine the exact point between both visits when the symptomatology has appeared. 
\end{itemize}

This paper focuses on right censoring, which is the most common censoring case in clinical studies \cite{Prinja}. Therefore, in order to deal with censoring, some specific notation needs to be introduced \cite{Klein2}. The random variable ``real lifetime", even though it is not always observed due to censoring, is represented by $Y$. Its probability density function is denoted by $f(y)$ and its survival function by $S(y)$. 

\paragraph{}
The maximum lifetime that can be observed for each individual, due to the aforementioned limitations, is known as ``censoring time" and denoted by $C$. Its value is determined by the end of the study, or by the moment of the loss of follow-up, and can be different for each individual. Therefore, the variable $Y$ can be observed only if condition $Y \leq C$ is fulfilled. Otherwise, the observation is censored and only $C$ is observed. The random variable ``observed lifetime", $T$, is defined as $T=\min(C, Y)$. If the observation is not censored, the observed lifetime will be equal to the real lifetime. Moreover, $\delta$ is the uncensoring indicator:
\begin{equation}
\label{eq:4}
\delta = \textbf{1}(Y \leq C).
\end{equation}

As defined by (\ref{eq:4}), $\delta$ is equal to $0$ if the observation is censored, and it is equal to $1$ otherwise. The sample is just a collection of independent observations $(T_1, \delta_1), \ldots , (T_n, \delta_n)$ with the same distribution as the random variables ($T$, $\delta$). 

\paragraph{}
Data analysis tools for the survival curve consider censoring. One of the most popular estimators is the Kaplan-Meier (KM) estimator by \cite{KaplanMeier}, named after the researchers who developed it. The KM estimator is a nonparametric estimator, and thus it does not make any assumption with regards to the specific form of the probability distribution of $Y$. This method considers that the probability of surviving beyond time $t$ from the beginning of the study equals the product of the $n$ survival rates in the period $[0,t]$ \cite{Bewick}. Mathematically, it is defined as: 

\begin{equation}
\label{eq:5}
\hat S(t)= \prod_{i:T_{(i)}\leq t}^n \left ( 1 - \frac{\delta_{[i]}}{n-i+1} \right ).  
\end{equation}

In (\ref {eq:5}), $\delta_{[i]}$ is the corresponding uncensoring indicator concomitant of $T_{(i)}$, and $T_{(1)} \leq T_{(2)} \leq \dots \leq T_{(n)}$ are the ordered observed lifetimes. It has been proved that this is the nonparametric maximum likelihood estimator for $S(t)$ \cite{Johansen}. Furthermore, a generalization of KM has been proposed in such a way that it allows considering the effect of a certain covariate $X$ in $S(t)$. This generalization, introduced by \cite{Beran}, is known as the Beran estimator, defined as: 

\begin{equation}
\label{eq:6}
\hat S_h(t|x) = \prod_{i:T_{(i)}\leq t} \left ( 1 - \frac{\delta_{[i]} B_{h(i)}(x)}{\sum_{r=i}^n B_{h(r)}(x)}  \right  ),
\end{equation}
where
\begin{equation}
\label{eq:7}
B_{h(i)}(x)= \frac{K_h(x-X_{[i]})}{\sum_{j=1}^n (K_h(x-X_{[j]}))}.
\end{equation}
In (\ref{eq:7}), $K_h$ represents the rescaled kernel function with a smoothing parameter, $h$, and $X_{[i]}$ is the covariate concomitant of $T_{(i)}$. A suitable choice of $h$ is critical in kernel estimation, and several bandwidth selection methods can be considered, such as bootstrap \cite{Lopezetal1}, plug-in \cite{Dabrowska1}, or cross-validation \cite{Iglesias-Perez2}. 

\subsection{Survival analysis versus cure models}
\label{sec:3}
Classical survival analysis assumes that all the individuals will experience the event of interest. However, in some cases, a fraction of the population will never experience failure. This fact is especially relevant in cancer studies: when the event of interest is the recurrence of a tumor and the study variable is the ``time from remission to relapse", some individuals will never suffer from a new tumor, even after a long time period. These individuals are said to be cured. Note that in a general survival analysis framework \emph{cure} differs from the classical meaning of absence of illness, but refers to the fact that these patients are exempt to suffer the event of interest. 

The application of cure models has not been limited to Medical Sciences or Epidemiology, but has also been extended to different areas such as Social Sciences, Economy or Engineering. For instance, the elapsed time from marriage to divorce, the lifespan of a specific product, or the unemployment period until an individual gets a job are all considered as time-to-event studies \cite{Emmert}, \cite{Ciuca}.

\paragraph{}
From a practical point of view, distinguishing between cured individuals and censored observations which are susceptible to experience the event of interest is not trivial. Cure models handle this situation, becoming essential statistical techniques in cases where applying a classical survival analysis is not appropriate.

\paragraph{}
The aim of this work is to review the distinct types of cure models, comparing their strengths and limitations. Furthermore, the different approaches will be applied to a COVID-19 dataset, studying the elapsed times from the diagnosis until hospitalization. 

\section{Cure models}
\label{sec:4}
The main idea of cure models is the willingness to complete the unavailable information in order to identify and estimate the fraction of cured individuals \cite{Patilea}. In this case, the survival function for the population does not fulfill one of the aforementioned assumptions which hold in classical survival analysis, since $\lim_{t \rightarrow \infty} S(t)>0$. The value of this limit is denoted as $1-p$ and corresponds to the proportion of cured individuals in the population or cure rate \cite{Amico1}. Estimating the value of $1-p$ is one of the main objectives of these models.

\paragraph{}
From their first appearance in 1949, various types of cure models have been proposed, which can be classified into two main groups: mixture cure models and promotion time cure models. The latter were designed as biological models for the analysis of relapse times in cancer studies \cite{Lambert}. They were formally proposed by \cite{Yakovlev2}, and they were initially used to model the tumour latency \cite{Yakovlev1}. In such a case, it is assumed that after the first diagnosis and successful treatment, a number $N \geq 0$ of carcinogenic cells remain in the organism in a latent form, each one of them for a period of time $T_k$, until they finally develop a new tumour. Those individuals for whom $N \geq 1$ present at least one carcinogenic cell and are susceptible to relapse, whereas those with $N=0$ are said to be cured and the latency time $T$ is infinite \cite{Amico1}. 
\paragraph{}
Promotion time cure models assume that $N$ follows a Poisson distribution with parameter $\theta>0$, which is the average number of carcinogenic cells in the population. Assuming that the different $T$ are i.i.d. with probability distribution $F(t)$, and independent from $N$, it can be demonstrated that the survival function of the population is defined by
\begin{equation*}
S(t)=\exp[-\theta F(t)].
\end{equation*}
\paragraph{}
Mixture cure models are a sort of two-part models which were firstly introduced by \cite{Boag}. These models study the response variables in two separate groups, which are identified by a binary variable, $B$, which is equal to 0 if the individual belongs to the cured group, and it is equal to 1 if the subject is susceptible to suffer the event of interest. Therefore, $B$ is the indicator variable for the susceptibility, and it is only partially observed since it is not possible to distinguish between those susceptible observations that are censored and those observations of cured individuals. 
\paragraph{}
In the context of survival analysis with a fraction of cured individuals, mixture models define the survival function of the population as: 
\begin{equation}
\label{eq:9}
S_{pop}(t|\textbf{x,z})=1-p(\textbf{x})+p(\textbf{x})S_u(t|\textbf{z}).
\end{equation}
In (\ref{eq:9}), \textbf{X} and \textbf{Z} are two sets of covariates which might be equal or not, and $p(\textbf{x})=P(B=1 | \textbf{X}=\textbf{x})$ represents the probability of being susceptible given the value of the covariates, \textbf{X}, and it is known as the incidence of the model. On the other hand, $S_u(t|\textbf{z})=P(T>t| \textbf{Z}=\textbf{z},B=1)$ is the survival function for the susceptible group, conditioned to the set of covariates, $\textbf{Z}$, and it is known as the latency of the model \cite{Amico1}. The model formulation provides that the cure rate, $1-p(\textbf{x})$, depends only on $\textbf{X}$, whereas the survival function of the susceptible group depends only on $\textbf{Z}$. The fact of having these two groups of covariates separated for cured and uncured individuals allows us to consider external factors to have different influence in both groups of patients. This is the main advantage of mixture cure models, the methodology in which this paper is focused. Depending on the assumptions established for the latency and the incidence of the model, there are parametric, semiparametric and nonparametric approaches of cure models. 

\paragraph{Parametric models}
Parametric mixture cure models were the first mixture cure models developed, and can be considered the basis of any further research in this field during the past 50 years. They were introduced by \cite{Boag} to study a mouth cancer cohort who had been treated with a certain therapy, with the intention to model the relapse of these patients. \cite{Boag} considered the cure rate as constant, and the survival function of the susceptible individuals was modeled according to a lognormal distribution with independence to any external covariate. \cite{Berkson} studied deeply the approach from \cite{Boag}, considering an exponential model for the latency, also with independence of covariates. \\

More than 30 years after the original proposal, the first cure models considering the influence of covariates were developed. \cite{Farewell1} proposed a new model where the latency followed a Weibull distribution, dependent on a set the covariates, \textbf{Z}, and modelling the incidence with a logistic function.  \\

These parametric models show limited flexibility, due to the strict assumptions with respect to the latency and the incidence distribution. A generalization of the latter models, but still maintaining the parametric behavior, are the models based on the accelerated failure time (AFT). They assume the presence of covariates where their effects are fixed and multiplicative by the accelerated factor on the time scale \cite{Saikia}:
 \begin{equation}
\label{eq:10}
\log(T^*)=\beta_0+\beta\textbf{Z}+\sigma\varepsilon.
\end{equation}
In (\ref{eq:10}), $T^*$ is the survival time of the susceptible individuals, and $\sigma$ is a scaling positive parameter. These models consider an error term ($\varepsilon$), whose density function is previously defined. AFT models were firstly proposed by \cite{Yamaguchi} and later developed by \cite{Peng2}. Note that the aforementioned \cite{Boag}, \cite{Berkson} and \cite{Farewell1} models can be derived from them, by giving specific values to the model parameters. \\

In all the cases, the estimation of the model parameters of parametric mixture models is performed using the maximum likelihood criterion, which is derived from classical survival models. The likelihood function for these models is defined as the product of two contributions: on the one hand, the censored observations and, on the other hand, the uncensored observations. It is not possible to distinguish between cured and uncured individuals in the censored part of the sample. 

\paragraph{Semiparametric models}
Semiparametric models arise from the necessity to improve the flexibility of the aforementioned approaches in order to extract information from the sample to a greater extent. They are said to be semiparametric since this flexibility is usually assigned to the latency, whereas the incidence is still modeled using parametric methods - usually, assuming a logistic regression for $p(\textbf{x})$. Depending on the assumptions made on the survival function of the susceptible group, we may find several types of semiparametric cure models: 
\begin{itemize}
\item Proportional hazards (PH) cure models. These models are based on the regression model, which is applied to general survival studies in order to model the risks that may affect a population. Therefore, these models are based on the hazard function, which is directly related to the survival function. In PH models, the hazard function of an individual is given by the product between the baseline function and a non-negative function of the covariates:
\begin{equation*}
\label{eq:11}
h(t|\mathbf{z})=h_0(t)c({\mathbf{\boldsymbol \beta^{\prime} z}}).
\end{equation*}
The function $c(\cdot)$ is known as the \emph{link function} and is frequently chosen as the exponential function \cite{Klein1}. Besides, $h_0(t)$ is the baseline hazard function and may take any possible shape, being parametric or nonparametric. Given the relationship between the hazard and the survival functions, the model can be expressed in terms of this latter function:
\begin{equation*}
\label{eq:11}
S(t|\textbf{z})=S_0(t)^{c({\mathbf{\boldsymbol \beta^{\prime} z}})},
\end{equation*}
where $S_0$ represents the survival baseline, defined as $S_0(t)=P(T>t|\textbf{Z}=0, B=1)$. These models were introduced by \cite{Kuk}, who adapted the parameter estimation methods for classical survival analysis to the presence of a cure fraction. Due to the additional condition $B=1$ in the definition of $S_0$, the traditional estimation methods cannot be applied, and thus additional tools were developed, based on expectation-maximization (EM) methods \cite{Amico1}.
\item AFT models. These models consist of a semiparametric adaptation of the aforementioned AFT models for the latency. These were introduced by \cite{Li1}, allowing the error term to present any survival functions without restrictions. Similarly as in the PH models, the incidence is modeled using a logistic function. The parameter estimation is performed using the maximum likelihood criterion, via EM methods. 
\item Flexible models. The aforementioned semiparametric approaches considered a logistic regression for incidence estimation. Nonetheless, \cite{Lam} proposed an enhancement in the model flexibility by introducing new functions to infer the cure rate, such as the probit or log-log distributions. They also considered the EM algorithm and the maximum likelihood criterion to estimate the parameters. \\ This flexibility, however, might not be enough to gather all the information within the study since incidence estimation is still parametric. Although parametric methods show some important advantages, such as their ease of interpretation or the simplicity of the parameter estimation, some authors have proposed semiparametric approaches to estimate the cure rate. Some of these semiparametric approaches are based on splines \cite{Wang} or on single-index structures \cite{Amico2}. The flexibility can be further improved if completely nonparametric forms are introduced for the latency, being independent of any external covariate \cite{Taylor} or even taking these external factors into account \cite{Patilea}, while the incidence is still modeled using parametric formulations. 
\end{itemize}

\paragraph{Nonparametric models}
In the aforementioned cases, at least one of the model components, namely the latency or the incidence, was defined by a (semi-)parametric formulation. However, a completely nonparametric approach for both elements can be also considered, achieving thus the maximum flexibility of the models. The first nonparametric models were proposed by \cite{Maller1}, but the authors did not consider the influence of covariates in the latency and incidence. This issue was partially solved by \cite{Laska}, who considered discrete variables. \\ The main progress in this field was carried out by \cite{Lopezetal1}, who proposed a completely nonparametric model, based on the nonparametric estimator for the cure rate developed by  \cite{Xu}. This estimator is also based on the Beran estimator for the survival function (see (\ref{eq:6})). Regarding the selection of the smoothing parameters in this nonparametric context, \cite{Lopezetal2} introduce a bootstrap bandwidth selection method for both the latency and the cure rate estimations. Furthermore, \cite{Lopezetal3} proposed a nonparametric covariate hypothesis test for the incidence in mixture cure models, which can be applied to continuous, discrete and qualitative variables. This test allows for the identification of those variables that play a significant role on the cure rate. This nonparametric model does not assume any previous restriction, and therefore, it can be completely adjusted to the data.

\section{Mixture cure models applied to COVID data}
\label{sec:5}
\subsection{Dataset}
\label{sec:6}
A practical study has been performed using a COVID-19 database to illustrate the application of cure models. COVID-19 databases are broadly studied in the present time, as reviewed in \cite{Shuja}, to develop a variety of mathematical models on the disease features \cite{Mohamadou}.  The data was extracted from the \textit{Servicio Galego de Sa\'ude} and provided by the \textit{Direcci\'on Xeral de Sa\'ude P\'ublica} (Galicia, NW Spain). The dataset consists of 4307 COVID-19 patients who tested positive by PCR, being thus a representative cohort which has also been used to model several features related to COVID-19 disease, such as the disease severity \cite{Gude} and the hospital and intensive care unit (ICU) length-of-stay \cite{Lopezetal4}. The available variables for each individual are:
\begin{itemize}
\item ID number of the patient, unique and anonymous
\item The age of the individual at diagnosis
\item The sex (male/female) of the patient
\item The date when the PCR test was first performed 
\item The hospital admission date, in case it was necessary due to the severity of the symptoms
\end{itemize}
The variable of interest is defined as ``time from diagnosis to hospital admission". Hospital admission and bed occupation is an important issue that needs to be addressed in order to cope with the current situation, and thus it has been the target of a large number of investigations \cite{Lapidus}, \cite{RLi}, \cite{Moghadas},  \cite{XQi}, \cite{Rees}, \cite{QThai}, \cite{ZWang}, \cite{Wood} . Since this is a time-to-event variable, survival methods can be used, as \cite{Prieto} did in their investigations concerning this variable for a Catalonian cohort \cite{Prieto}. Furthermore, there is a large fraction of individuals in the dataset (around 90$\%$) who had not been admitted into hospital at the end of the follow-up period. Part of these individuals might be censored observations and thus, admission would occur after the end of the study. However, there will also be a number of individuals which do not require hospitalization during the illness, being those cured observations. The dataset features therefore justify the application of the cure models presented above.
\paragraph{}
The database analysis by means of survival and cure models was performed using R software \cite{Cai} \cite{Ullibarri}.

\subsection{Preliminary analysis}
\label{sec:7}
\begin{figure*}

  \includegraphics[width=0.75\textwidth]{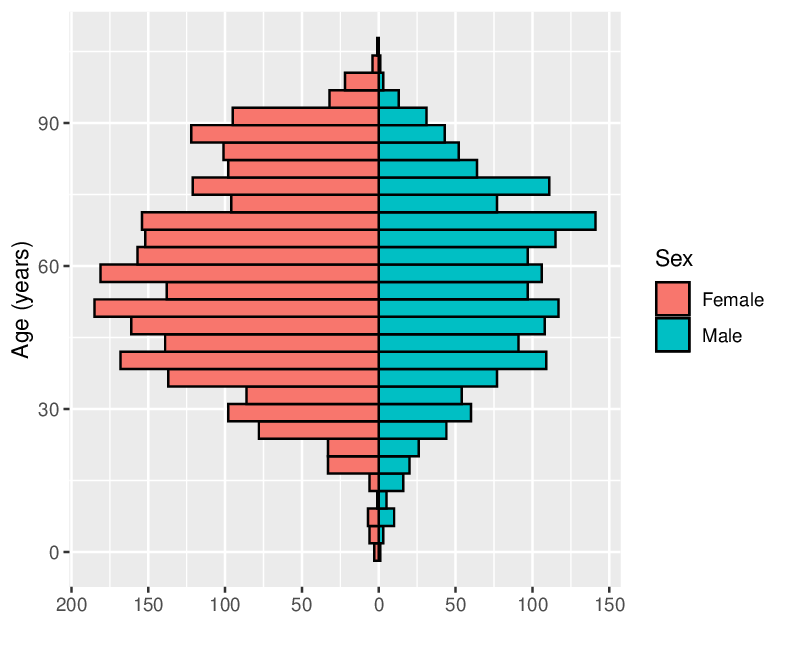}
\caption{Demographic distribution of the sample in terms of age and sex.}
\label{fig:1}       
\end{figure*}
\begin{figure*}
  \includegraphics[width=0.75\textwidth]{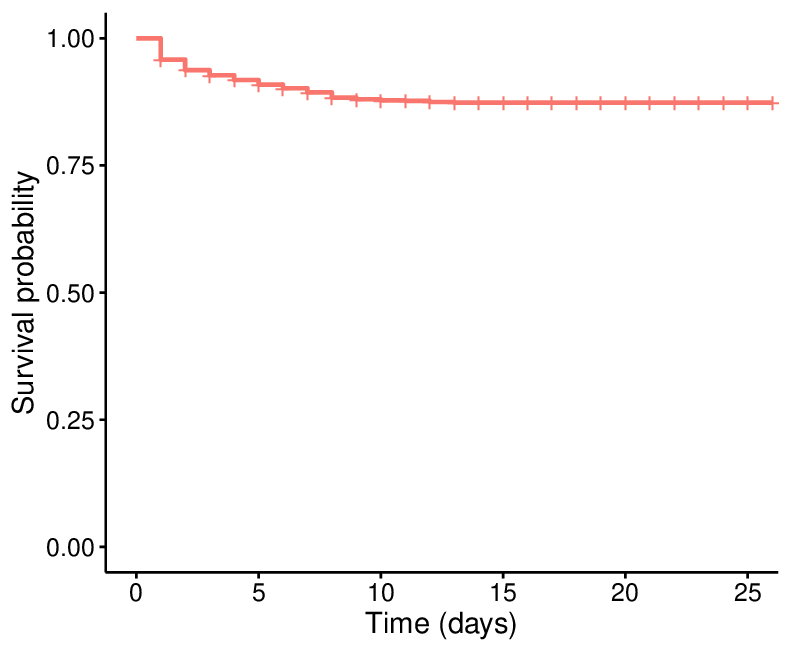}
\caption{Survival function estimation using the KM method for time to hospital admission.}
\label{fig:2}       
\end{figure*}

From the $4307$ patients of the sample, $2615$ ($60.72\%$) are female. The average age is $57.1$ years. The average age for men is $56.3$ years and for women is $57.6$ years. The distribution of the sample in terms of age and sex is represented in Fig. \ref{fig:1}.
\paragraph{}
Data registration started on March, $6^{{\rm th}}$, 2020, and the first hospitalization occurred a day after. The last diagnosis observation is placed on April, $2^{{\rm nd}}$, whereas the last observed hospitalization took place on April, $3^{{\rm rd}}$. The minimum survival time, defined as the elapsed time from diagnosis until hospital admission is $1$ day, being $13$ days the maximum uncensored observation of the time-to-event variable. 
\paragraph{}
A first insight of the study variable behavior can be performed using the KM estimator for the survival curve (Fig.~\ref{fig:2}). By using this approach, it is possible to assess some of the main features of the dataset. The points marked with the $+$ symbol correspond to censored observations, which are distributed all over the time line. The jumps represent the time points where the failure has occurred for the uncensored observations. The form of the curve is due to the characteristics of the individuals in the database: the diagnosis and hospitalization times were registered in a short date format, considering only the day of diagnosis. Thus, the survival times are discrete observations and, considering the relatively short extension of the study, there exist large leaps that finally lead to the observed curve shape. Furthermore, we can also observe a \textit{plateau} in the curve, which extends from the last uncensored observation on. 
\paragraph{}

\begin{figure*}
  \includegraphics[width=0.5\textwidth]{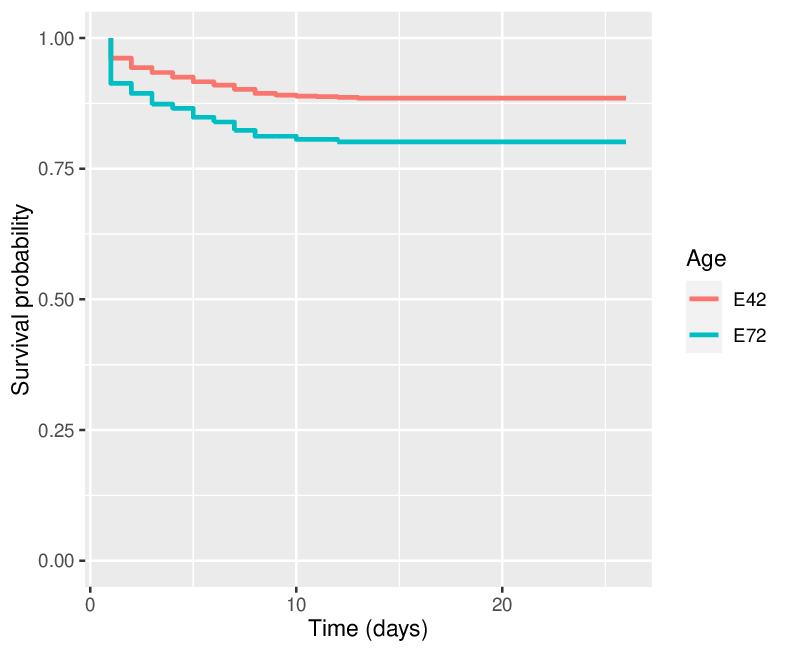}
\includegraphics[width=0.5\textwidth]{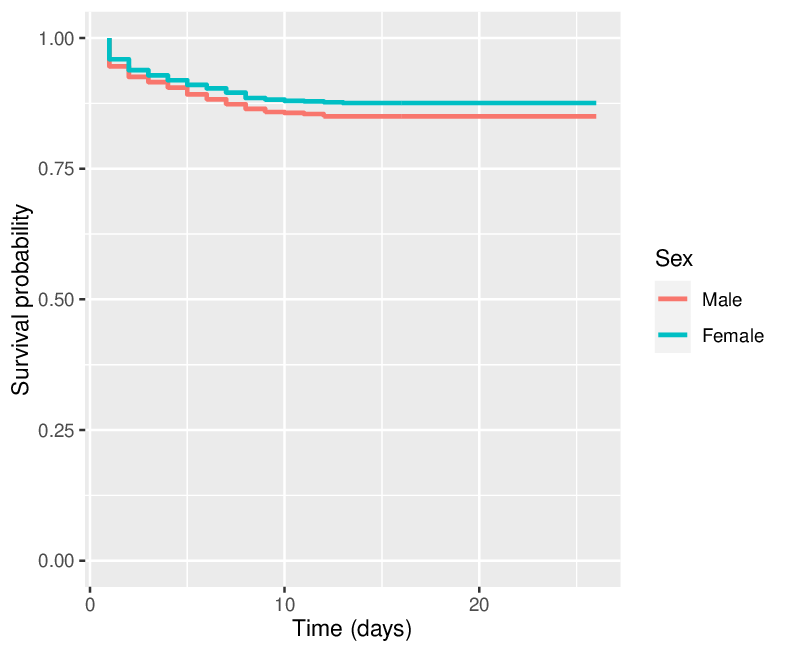}
\caption{Estimated survival curves obtained by the Beran method, which considers the survival conditioned to a set of covariates. On the left, the survival curve estimated for two individuals aged 42 and 72 years, aiming to illustrate the influence of this variable. On the right, the estimated survival curves of women and men, which allow us to observe the influence of the sex of the patient on the study. }
\label{fig:3}       
\end{figure*}
In the same way, a preliminary analysis for the covariate influence (age and sex of the patient) can also be performed by using the Beran estimator presented in (\ref{eq:6}). Specifically, the survival curves conditional on the age and sex were calculated, and the shape of these curves was observed for several values of these covariates. In Fig.~\ref{fig:3} a comparison of the curves obtained with different covariate values is presented, when the ``sex" variable is equal to 0 (male) or 1 (female), or when the covariate ``age" is equal to the $1^{{\rm st}}$ or $3^{{\rm rd}}$ sample quantiles, namely $42$ and $72$ years. Apparently, the age covariate has an increased influence in the survival when compared to the covariate ``sex", since the conditional survival functions are more far apart. \\

In order to obtain further information about the behavior of the study variable, it is necessary to apply the aforementioned cure models. Therefore, in this study we will apply semiparametric approaches, implemented in the \textit{smcure} R package by \cite{Cai}, and nonparametric approaches, using functions from the \textit{npcure} R package by \cite{Ullibarri}. 

\subsection{Semiparametric cure models }
\label{sec:7}
The \textit{smcure} package contains the tools needed to apply mixture cure models to a given dataset in a semiparametric context \cite{Cai}. It considers Cox PH-derived models and AFT-derived models, as those presented in Section~\ref{sec:4}. The implementation of the \emph{smcure} function allows us to explicitly select the model, as well as to select the type of parametric regression for the incidence between the three following distributions: logit, probit and complementary-loglog (cloglog). The parameter adjustment is performed using the EM algorithm under the maximum likelihood criterion. 
\paragraph{}
A total of $6$ different semiparametric mixture cure models were defined by combining the two available models for the latency (PH and AFT models) with the three possibilities for the incidence regression (logit, probit and cloglog). In all the models, both the sex and the age of the patient were considered as covariates for both the latency and the incidence. The significance of these covariates obtained for each of the models is presented in Tables~\ref{tab:1} and ~\ref{tab:2}. The covariates could not be considered as significative in any of the cases, and thus we cannot claim that the age or the sex of an individual affects the probability of needing hospitalization or the time since diagnosis until hospital admission. 

\begin{table}
\caption{Significance of the covariates ``sex" and ``age" on the latency of the proposed semiparametric models.}
\label{tab:1}      

\begin{tabular}{llll}
\hline\noalign{\smallskip}
Latency model & Variable & Coefficient estimation & p-value  \\
\noalign{\smallskip}\hline\noalign{\smallskip}
proportional hazards Cox model & age & $1.26 \cdot 10^{-3}$ &0.641\\
proportional hazards Cox model & sex & $2.63 \cdot10^{-2}$ &0.744 \\
accelerated failure time & age & $0.00$ &1.00 \\
accelerated failure time & sex & $0.00$ &1.00 \\
\noalign{\smallskip}\hline
\end{tabular}
\end{table}

\begin{table}
\caption{Significance of the covariates ``sex" and ``age" on the incidence for the three approaches considered.}
\label{tab:2}      

\begin{tabular}{llll}
\hline\noalign{\smallskip}
Incidence model & Variable & Coefficient estimation & p-value  \\
\noalign{\smallskip}\hline\noalign{\smallskip}
logit & age & $4.02 \cdot10^{-8}$ &0.186\\
logit & sex & $-1.66 \cdot10^{-7}$ &0.934 \\
probit & age & $-2.95 \cdot10^{-11}$ &0.999 \\
probit & sex & $4.51 \cdot10^{-10}$ &0.999 \\
cloglog & age & $-3.04 \cdot10^{-11}$ &0.932 \\
cloglog & sex & $2.42 \cdot10^{-9}$ &0.908 \\
\noalign{\smallskip}\hline
\end{tabular}
\end{table}

\begin{figure*}

  \includegraphics[width=0.75\textwidth]{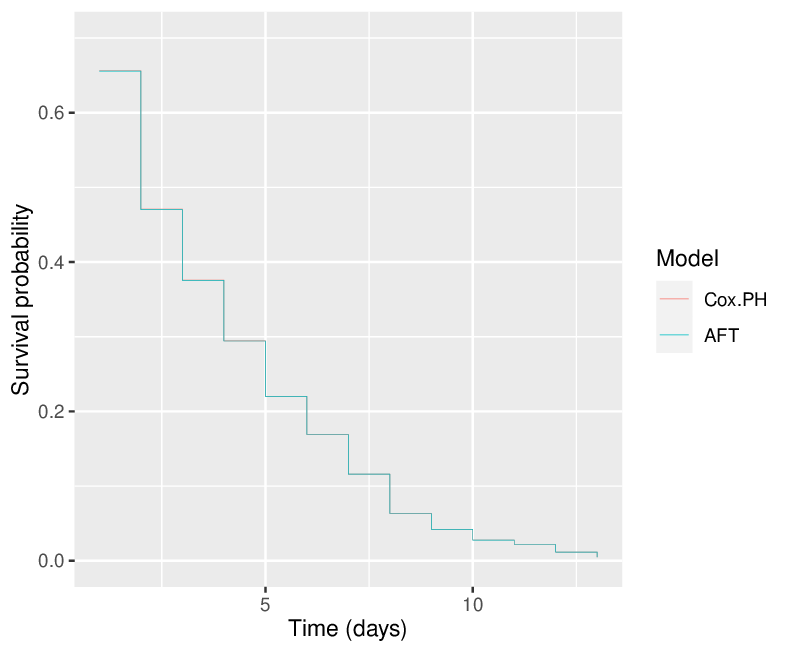}
\caption{Comparison between the latency estimated with both AFT and PH models for the same individual. The survival probability values are so close between models that is nearly impossible to distinguish both curves.}
\label{fig:4}       
\end{figure*}

The AFT and Cox PH models were compared by setting the sex and age covariates to some representative values in the sample. In this case, the survival curves were estimated considering a female individual, since females represent the majority of the sample, $57$ years old, which is the median age within the sample. The survival curves for this representative individual estimated with both latency models (AFT and PH), applying a probit regression for the incidence, are presented in Fig.~\ref{fig:4}. As we can observe, there is not a significant difference between both models. This can be due to the intrinsic limitations in the dataset used for our analyses. In this particular case, we can anticipate, on the one hand, the presence of a low fraction of susceptible individuals, since about $90\%$ of the observations are censored. On the other hand, the time-to-event variable is discretized, since the starting point considered is the day when the diagnosis took place, ignoring the exact moment within it. Even though there are some mixture cure models that specifically consider discrete time variables \cite{Zhao}, they have not been implemented in R yet.

\begin{figure*}

  \includegraphics[width=0.5\textwidth]{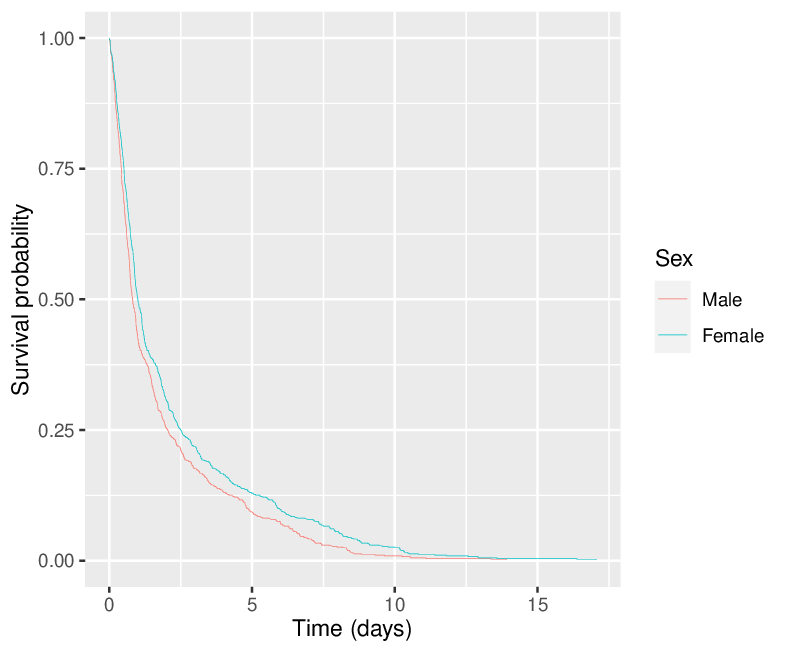}
  \includegraphics[width=0.5\textwidth]{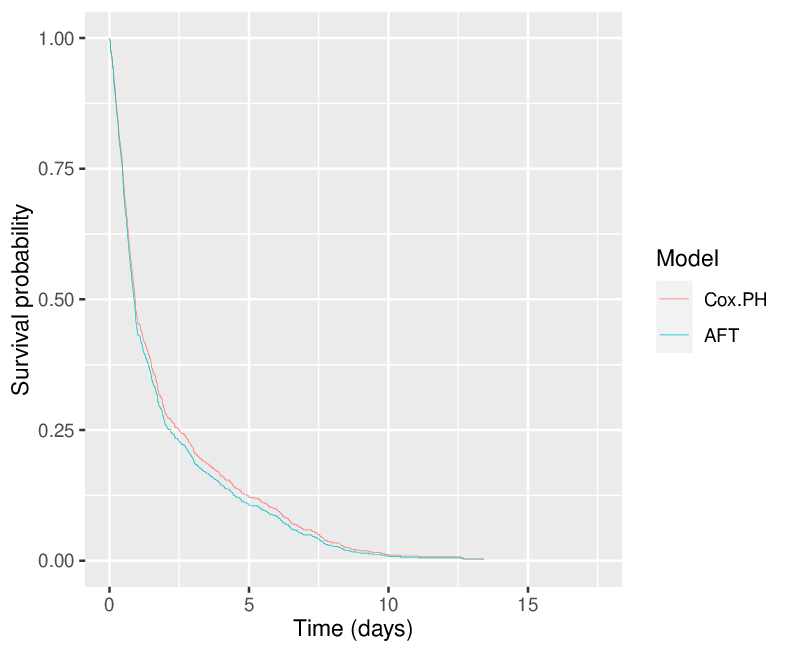}
\caption{Results obtained with the semiparametric models when working with continuous-time variables, which were calculated using the probit model for the incidence. On the left, the survival curves of susceptible patients are estimated separately for women and men using the AFT model. On the right, the survival curves are estimated for a 57 year-old female individual considering the two available semiparametric approaches.}
\label{fig:5}       
\end{figure*}

\paragraph{}
In order to tackle with the aforementioned limitations, we simulated the behavior of a continuous-time variable using as a basis the original dataset. For this, we perturbed the data with random variables derived from a uniform distribution $U(-1,1)$. As a result, a value between $0$ and $1$ is randomly subtracted or added to the original value of the time variable, and therefore the lifetimes will no longer be represented by integers. 
\paragraph{}
After applying this procedure and fitting again the previously presented models, it was found that the covariate ``sex" presented a significant influence on the latency of the AFT models, with a $p$-value $0.015$. This fact is also evident when representing the estimated survival curves for women and men, where we can notice that the survival of women is higher when compared to men. Thus, we can conclude that the period from the diagnosis until the hospital admission for those patients with severe symptomatology will be longer for women. The variable ``sex", however, was not found significant for the incidence of the model, concluding that with this approach men and women show equal probabilities of needing hospitalization. This reinforces the ideas by \cite{Gebhard} and \cite{Jian-Min}. However, recent researches on COVID-19 cohorts have shown that male patients exhibit a worse prognosis compared to women, which may lead to an early need for hospitalization, as shown by these results \cite{Jian-Min}. 
\paragraph{}
Furthermore, this analysis allows proper comparison between AFT and PH models. Besides the different results obtained for each of the models, in Fig.~\ref{fig:5} (right), we can see that for a $57$ years old female patient, a slightly different behavior between models is appreciated. This difference reinforces the importance of an accurate selection of the model characteristics in order to capture the information within the sample. 

\subsection{Nonparametric cure models}
The \textit{npcure} package by \cite{Ullibarri} can be used to apply nonparametric estimations for mixture cure models. It includes the methods developed by \cite{Lopezetal1}, \cite{Lopezetal2} and \cite{Lopezetal3}, which have been previously presented in Section~\ref{sec:4}. 
\paragraph{}
Besides the appropriate tools needed to study the latency and the incidence of the population for a certain event of interest, this package includes some covariate hypothesis tests that can be used to confirm the inclusion of a certain covariate in the incidence in a mixture cure model. 
\paragraph{}
Furthermore, the test by \cite{Maller1} is also implemented, which allows us to check if there is a cured fraction of individuals in the sample, and thus the suitability of cure models. Accepting or rejecting the null hypothesis determines the absence or presence of a pleateau in the survival curve, respectively. This plateau corresponds exclusively to the observations of cured individuals. The Maller-Zhou test was applied to the dataset, and a $p$-value equal to $0$ was found. Therefore, we can reject the null hypothesis and thus confirm that there exists a cured fraction of individuals within the sample. 
\paragraph{}
Another hypothesis test that can be used to study the significance of covariates for the cure probability is also included in the \textit{npcure} package. This test is based on \cite{Delgado}, and it was extended by \cite{Lopezetal3}, making it possible to determine whether the cure probability, is dependent of a given covariate $X$ or not:
\begin{equation*}
\left \{ \begin{matrix} H_0: \mbox{cure probability} &=& 1-p \\ 
H_1: \mbox{cure probability} &=& 1-p(x) \end{matrix}\right.  
\end{equation*}

These hypotheses were tested for the COVID-19 database, and the results show that, with a significance level $\alpha=0.01$, both sex and age influence significantly on the cure probability of the population, with $p-$values equal to $0.002$ and $0$, respectively. Thereby, we justify the inclusion of both covariates in our analyses.

\begin{figure*}
  \includegraphics[width=0.5\textwidth]{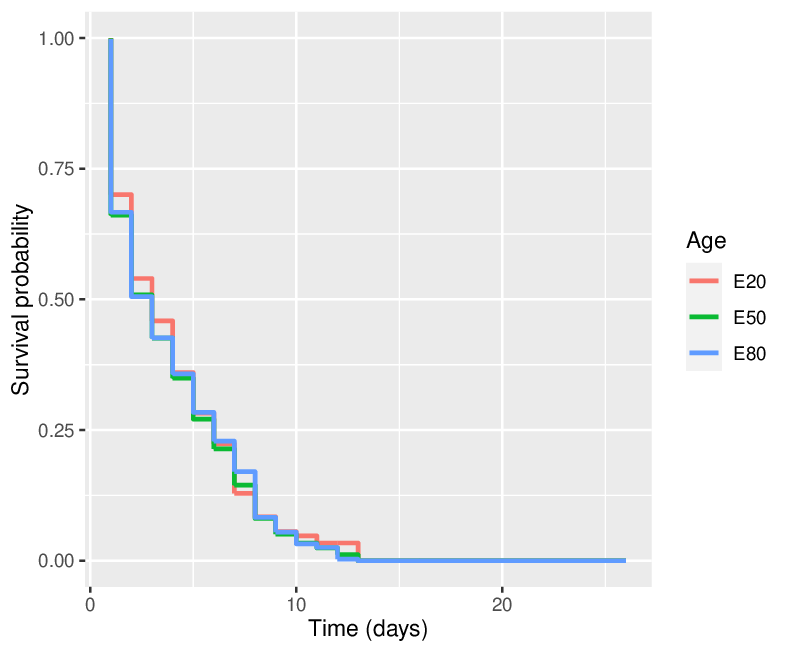}
\includegraphics[width=0.5\textwidth]{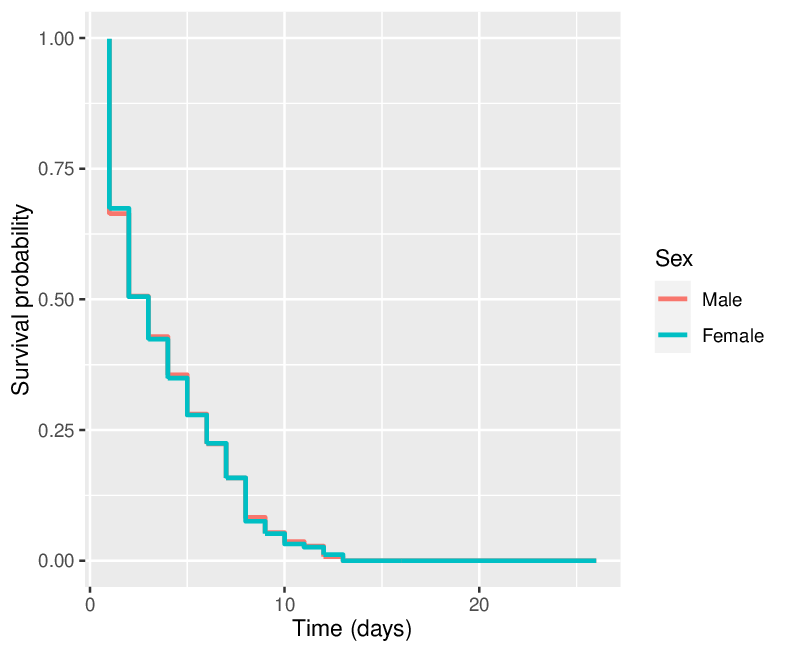}
\caption{Estimated survival function for the time until hospital admission of the susceptible patients using the nonparametric model, conditional on the covariates. On the left, model latency calculated for different values of the covariate ``age of the patient". On the right, latency of the model conditional on the sex of the patient. }
\label{fig:6}       
\end{figure*}

\paragraph{}
In the \textit{npcure} package, the estimation of the latency is implemented following the method described in \cite{Lopezetal2}, which is based on the Beran estimator for the calculation of the survival curve of the susceptible individuals, conditionally on a certain set of covariates. For each one of the covariates, namely age, sex, or a combination of both, a different curve is obtained. Since this estimation uses a kernel smoothing method, selection of the smoothing parameter is performed. The bootstrap method is used to mimic the minimization of the Mean Integrated Squared Error (MISE) criterion. The value of the bootstrap MISE is approximated using Monte Carlo, based on 100 bootstrap resamples. 
\paragraph{}
In order to analyze the effect of the age on the latency, we estimated the survival curve when this covariate is equal to $20$, $50$ and $80$ years, using the previously computed bootstrap smoothing parameter selector. The result is presented in the left part of Fig.~\ref{fig:6}. As it was expected, among the COVID-19 patients who needed hospitalization, those in their early stages of life tend to need it later after their diagnosis when compared with the elder population. The probability of the need for hospital admission at the beginning of the disease increases in the case of older people, and it is at that point in the course of the disease when the differences are more evident. This is consistent with the results of the epidemiological research that has been carried out in the last months, which claims that age is a clear risk factor for COVID-19 bad prognosis. Recently, it has been empirically observed that patients aged over $65$ are prone to the need for ICU admission or respiratory support, at the same time they present a decreased lymphocyte count compared to young individuals \cite{Richardson}. 
\paragraph{}
As for sex, even though the covariate hypothesis test showed that this factor implied a significant influence on the cure rate (that is, no need of hospital admission), the estimated survival function of male and female were practically equivalent, as it can be observed in the right part of Fig.~\ref{fig:6}. 

\begin{figure*}

\includegraphics[width=0.75\textwidth]{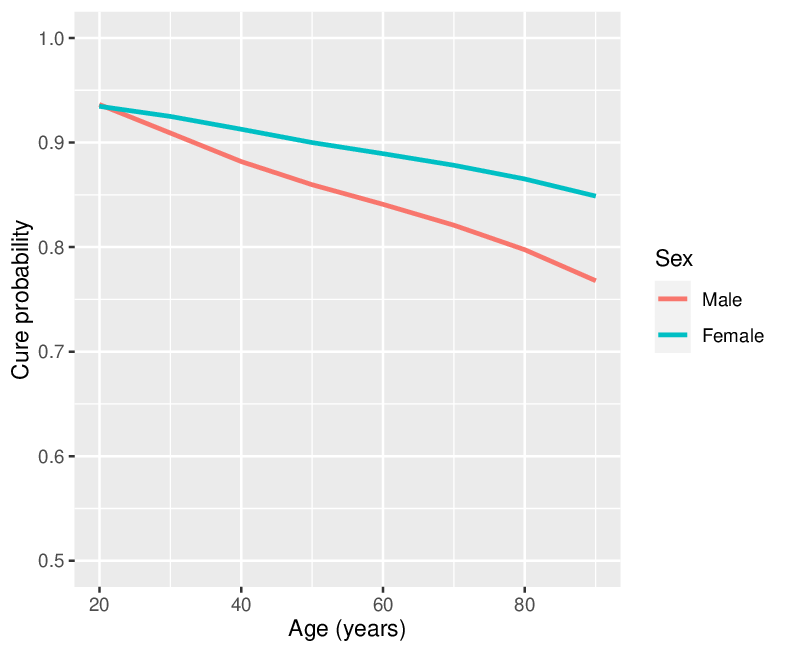}
\caption{Estimation of the cure probability (no need of hospital care) of COVID-19 patients with respect to the age, which shows a clear decreasing trend for both men and women.}
\label{fig:7}       
\end{figure*}

\paragraph{}
The covariate age also showed a significant influence on the incidence of the model, as it was anticipated by the covariate test (Fig.~\ref{fig:7}). In order to observe the changes in the cure probability when increasing the age of the patient, its value was estimated with the nonparametric method proposed by \cite{Lopezetal1} for an age interval between $20$ and $90$ years old. The cure probability showed a decreasing tendency when increasing the age of the patient. This is also consistent with the previous results on COVID-19 literature, since it has been claimed that there is a clear increase in the hospitalization rates with advanced ages, being elder adults those who needed more frequently hospital care \cite{Garg}. 
\paragraph{}
The covariate sex has a small influence in the cure probability, even though the covariate test considered a significant effect of this factor. This probability was found to be $0.870$ for men and $0.877$ for women. Sex, thus, apparently causes a difference of only $1\%$ in the probability of hospitalization after a COVID-19 positive diagnosis. The difference between male and female patients, however, appeared to be significant when these probabilities were calculated in the interval of ages between $20$ and $90$ years (Fig.~\ref{fig:7}). It is important to highlight that males need, on average, hospital care more frequently than women, and this need clearly increases when considering elder patients. Once again, these results fall into line with the literature \cite{Jian-Min}: men tend to suffer a more severe symptomatology and thus, to need hospital admission more frequently than women. 

\begin{figure*}

\includegraphics[width=0.5\textwidth]{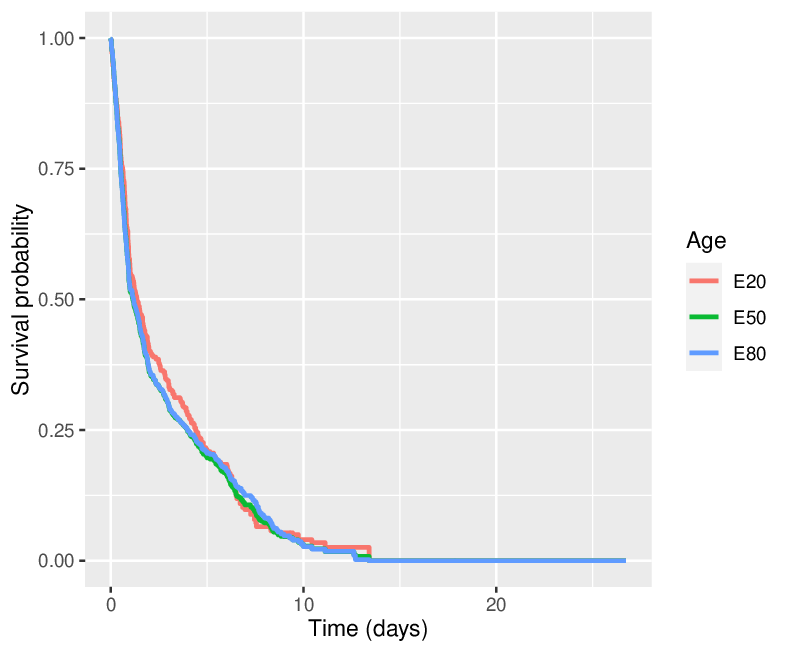}
\includegraphics[width=0.5\textwidth]{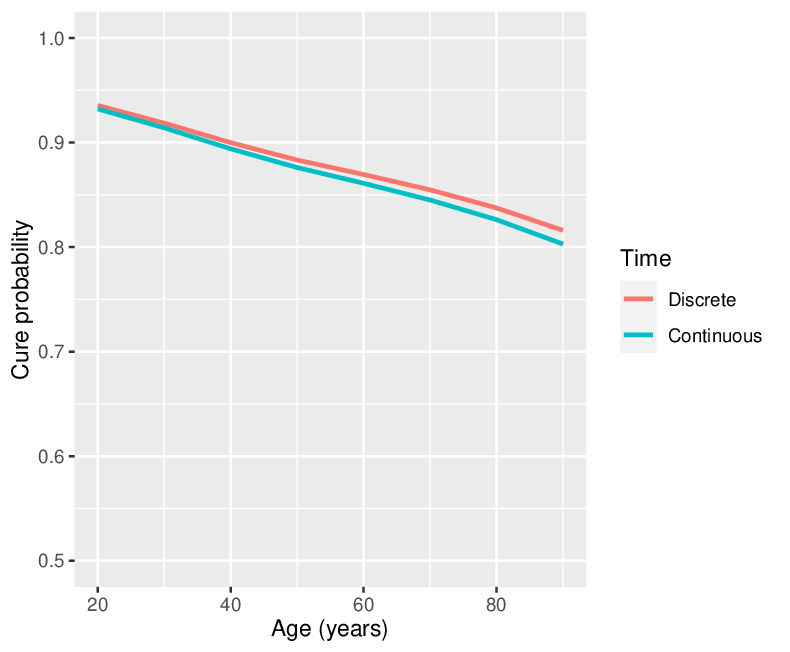}
\caption{Features of the nonparametric model when working with continuous data. On the left, comparison between survival curves of the patients with respect to their ages ($20$, $50$ and $80$). On the right, the trend of the cure probabilities when calculated with discrete and continuous times is compared.  }
\label{fig:8}       
\end{figure*}

\paragraph{}
In the case of nonparametric models, we also performed the data perturbation in order to obtain continuous-time data, following the same approach as with semiparametric models. The main difference when compared with the discrete data models was the loss of signification for sex in terms of cure probability ($p-$value=0.058), although the $p-$value obtained for this test is close to the significance level if we consider $\alpha=0.05$. The age is still found significant for such a  probability, and the estimated incidence is equivalent to the one obtained with the discretized times (Fig.~\ref{fig:8}, right side).

\paragraph{}
With regards to the latency, we obtained a smooth estimation of the survival curve when working with a continuous-time variable, which is more likely to be an accurate approximation of the real function. Furthermore, the differences between curves are easily observed when implementing the data perturbation. In the left side of Fig.~\ref{fig:8}, it can be noticed that the lifetime of young patients (that is, the elapsed time until their hospitalization), increases in the first stages of the disease, and then decreases until it is on the same level as the lifetime of the elder. This can be due to the current COVID-19 protocols: those patients over a certain age are directly admitted to the hospital when any minimal medical complication arises, whereas young individuals will only require hospital care if the symptoms are severe or if they last for a long time period. 

\section{Discussion}
\label{sec:9}
In our research, it has been claimed that different cure models lead to a variety of results and reach different conclusions, even though all of them belong to the same category of mixture cure models. When working with discrete lifetimes, nonparametric models apparently fit better to the actual situation, since there exist studies sustaining the influence of sex and age on COVID-19 prognosis. When using semiparametric models, we concluded that the set of covariates considered for this study does not yield any significant effect neither on their latency nor on their incidence, independently of the configuration selected for the model. Furthermore, it has been previously stated that these models might fail to reach convergence in the likelihood maximization, and thus resulting in biased estimators \cite{Lu}. Due to the intrinsic features of our data, showing some limitations such as the discretization of the times, the high rate of censored data and a foreseeable high cure probability, it is possible that semiparametric models are not a good choice for analyzing this data. Nonetheless, they have been previously used for other COVID-19 studies, specifically semiparametric Cox PH models, which also considered the influence of the age on the survival \cite{Sreedevi}. 
\paragraph{}
In order to tackle with some of the limitations, we analyzed the behavior of the models when working with continuous data. In this case, it was indeed possible to assess the effect of the sex of the patient on the latency. Therefore, the importance of working with exhaustive, high quality data in order to obtain feasible conclusions was reinforced.
\paragraph{}
Furthermore, nonparametric models were able to determine the effect of the study covariates on the time to hospital admission variable, even when it was discretized. With this approach, we studied the influence of the age on the survival of the population, and the results were consistent with the epidemiological data available in the literature. As far as we know, there are not previous publications considering completely nonparametric models to estimate the cure on COVID-19 cohorts, and this, together with the models developed by \cite{Lopezetal4} on the length-of-stay prediction, are pioneer on the field. 
\paragraph{} 
Cure models are not the only tool that can be used to extract information about COVID-19 patients. Indeed, most of the aspects concerning this disease have been tackled using different sorts of models and algorithms, including clinical de-identification of COVID-19 datasets \cite{Catelli} or forecasting of the number of future patients using ARIMA models \cite{HernandezMatamoros}. 

\paragraph{}
There also exist several studies aiming to predict the hospital and ICU admission based on different covariates such as  age, gender and medical conditions of the patients, using classification algorithms such as Support Vector Machines and Random Forest \cite{Pereira} \cite{Davila}. This leads to conclusions that can be completed with the results of this work, since this will give information of when that hospital admission will take place. However, machine learning-based approaches are not the best option when working with censored data, although they are specially helpful when handling high-dimensional clinical data \cite{Spooner}. Note that, in a context with censored data, it is not possible to apply directly machine learning classical models since they do not account for censored observations. Therefore, machine learning techniques should be adapted so that they also consider individuals who do not experience the event of interest. Thus, a few techniques have arisen in order to adapt this kind of predicting algorithms to the peculiarities of censored time-to-event observations, such as likelihood-based approaches \cite{Stajduhar} or the inverse probability of censoring \cite{Vock}. Nonetheless, these methods are not completely developed and, as far as we know, theoretical results that prove their good behavior have not been presented yet. For future research, it will be interesting to study and propose a complete adaptation of machine learning techniques to the context of censored data. Therefore, both approaches will contribute in the analysis, leading to a more accurate result.

\section{Conclusions}
Over the last 60 years, there has been a remarkable progress in cure models and thus several implementations of these tools are available. In this work, we pointed out the differences among them, firstly from a theoretical perspective and later by their application to a real dataset. It has been observed that different model implementations reach a variety of conclusions related to the same dataset, which highlights the importance of using a suitable model when studying time-to-event data. 
\paragraph{}
On the other hand, this paper emphasizes the importance of having high-quality datasets. It has been observed that working with discrete data leads to completely different results than considering continuous data. It is important to note that, besides choosing a suitable model for the data, using a representative and informative database is also an essential part in the analysis.
\paragraph{}
When working with proper variables and suitable models, it has been possible to analyze some key aspects of hospitalization in COVID-19 and its relationship with the variables sex and age of the patient. Therefore, it has been proved that cure models are helpful tools for this kind of studies.

\begin{acknowledgements}
MPL activity was funded by the Science, Technology, and Innovation Plan of the Principality of Asturias (Spain) Ref: FC-GRUPIN-IDI/2018/000225, which is part-funded by the
European Regional Development Fund (ERDF). ALC was sponsored
by the BEATRIZ GALINDO JUNIOR Spanish Grant from MICINN
(Ministerio de Ciencia, Innovación y Universidades) with reference BGP18/00154. RC and ALC acknowledge partial support by the MINECO grant MTM2017-82724-R, and by the Xunta de Galicia: Grupos de Referencia Competitiva ED431C-2020-14, Centro de Investigación del Sistema universitario de Galicia ED431G 2019/01, and Axencia Galega de Innovación (Ayudas proyectos de investigación COVID-19 presentados a la convocatoria del ISCIII IN845D 2020/26 - Programa Operativo FEDER Galicia 2014-2020), all of them through the ERDF.
\end{acknowledgements}

%
%

\bibliographystyle{spbasic}      
\bibliography{biblio}   


\end{document}